\documentclass[9.5pt,conference]{IEEEtran}
\usepackage{amsmath}
\usepackage{amsthm}
\usepackage{cite}
\usepackage{amssymb}
\usepackage{graphicx}
\usepackage{algorithm}
\usepackage{algorithmic}
\newtheorem{theorem}{Theorem}

\ifCLASSINFOpdf
\else
\fi
\begin{document}
\title{Input-Output Relation and Low-Complexity Receiver Design for CP-OTFS Systems with Doppler Squint}
\author{
	Xuehan~Wang\textsuperscript{1},~Xu~Shi\textsuperscript{1},~Jintao~Wang\textsuperscript{1,2},~Jian~Song\textsuperscript{1,2,3}\\
	\IEEEauthorblockA{
		\textsuperscript{1}Beijing National Research Center for Information Science and Technology (BNRist),\\
		Dept. of Electronic Engineering, Tsinghua University, Beijing, China\\
		\textsuperscript{2}Research Institute of Tsinghua University in Shenzhen, Shenzhen, China\\
		\textsuperscript{3}Shenzhen International Graduate School, Tsinghua University, Shenzhen, Guangdong, 518055\\
		\{wang-xh21@mails., shi-x19@mails., wangjintao@, jsong@\}tsinghua.edu.cn}
\thanks{This work was supported in part by Tsinghua University-China Mobile Research Institute Joint Innovation Center.}	
}
\maketitle
\begin{abstract} 
In orthogonal time frequency space (OTFS) systems, the impact of frequency-dependent Doppler which is referred to as the Doppler squint effect (DSE) is accumulated through longer duration, whose negligence has prevented OTFS systems from exploiting the performance superiority. In this paper, practical OFDM system using cyclic prefix time guard interval (CP-OFDM)-based OTFS systems with DSE are adopted. Cyclic prefix (CP) length is analyzed while the input-output relation considering DSE is derived. By deploying two prefix OFDM symbols, the channel estimation can be easily divided into three parts as delay detection, Doppler extraction and gain estimation. The linear equalization scheme is adopted taking the block diagonal property of the channel matrix into account, which completes the low-complexity receiver design. Simulation results confirm the significance of DSE and the considerable performance of the proposed low-complexity receiver scheme considering DSE. 
\end{abstract}

\IEEEpeerreviewmaketitle
\section{Introduction}
Reliable data transmission in high-mobility scenarios where the velocity approaches $1000$km/h is regarded as one of the key requirements in the 6G mobile network \cite{6G_velocity_ref}, where orthogonal time frequency space (OTFS) modulation is becoming a promising alternative to generate the communication waveform. By placing data symbols in delay-Doppler grids, the doubly-selective fading caused by the high-mobility and multipath propagation can be easily mitigated since full diversity over time-frequency is utilized for each symbol. So far, extensive work has verified the performance superiority of OTFS compared with traditional schemes such as orthogonal frequency-division multiplexing (OFDM) \cite{OTFS_performance1,OTFS_performance4}. \par
For practical OTFS systems, linear minimum mean squared error (LMMSE)-based equalizer can satisfactorily address the trade-off between the complexity and reliability \cite{OTFS_LMMSEdata2,OTFS_LMMSEdata1}. However, the attainment of channel state information (CSI) remains an open problem and has drawn substantial attention \cite{OTFS_CE_thres1,OTFS_CE_thres3,OTFS_CE_MP,OTFS_CE_EMVB}. Threshold-based delay-Doppler detection was developed, where the threshold can be fixed \cite{OTFS_CE_thres1} or adaptive \cite{OTFS_CE_thres3}. Meanwhile, compressed sensing-based schemes usually outperform the threshold-based methods by exploiting the sparsity of parameters, e.g., the Bayesian approaches \cite{OTFS_CE_MP,OTFS_CE_EMVB}.\par
However, the subcarrier-dependent Doppler which is referred to as the Doppler squint effect (DSE) exists and the deviation is accumulated through much longer duration in OTFS systems, which leads to severe performance degradation \cite{OTFS_DSE_mine} in most of existing schemes \cite{OTFS_performance1,OTFS_performance4,OTFS_LMMSEdata2,OTFS_LMMSEdata1,OTFS_CE_thres1,OTFS_CE_thres3,OTFS_CE_MP,OTFS_CE_EMVB,OTFS_cross_sigmodel,OTFS_cross_sigmodel2} due to the fundamental input-output analysis \cite{OTFS_cross_sigmodel,OTFS_cross_sigmodel2} ignoring DSE. To the best of the authors’ knowledge, \cite{OTFS_DSE_mine} is the only work considering DSE. However, impractical waveform and low transmission efficiency are involved, which inspires more efficient and practical design for OTFS systems with DSE. \par
In this paper, OFDM system using cyclic prefix time guard interval (CP-OFDM)-based OTFS is adopted. The input-output relation is analyzed based on the multipath linear time-variant (LTV) model with DSE, where the cyclic prefix (CP) length is elaborately designed to ensure no inter-symbol interference (ISI) between OFDM symbols within an OTFS symbol. The low-complexity receiver is then provided, where channel parameters are recovered by delay detection, Doppler extraction and gain estimation while the LMMSE-based equalizer is employed for each OFDM symbol. Finally, simulation results confirm the significance of DSE and the excellent performance of the proposed low-complexity receiver considering DSE. \par     
\textit{Notations}: $\mathbf{A}$, $\mathbf{a}$, $a$ denote a matrix, column vector and scalar, respectively. $\mathbf{A}^H$ and $\mathbf{A}^{-1}$ are its conjugate transposition and inverse. $||\mathbf{A}||_F$ denotes the Frobenius-norm of $\mathbf{A}$. $\lceil\cdot\rceil$ represents the ceiling function while $(\cdot)_{N}$ denotes the modulus operation with respect to $N$. $\overline{\mathbf{a}}$ returns the average value of $\mathbf{a}$. Finally, $\angle a$ returns the phase of complex value $a$.
\section{System Model}
CP-OFDM-based OTFS system \cite{OTFS_CE_EMVB} with the frame structure in Fig. \ref{Fig_frame} is investigated in this section. The multipath LTV channel with DSE is also presented, where we analyze why it cannot be ignored in OTFS systems. $f_{c}$ and $\Delta f$ denote the carrier frequency and subcarrier spacing, respectively. The impact of noise is disregarded for ease of illustration.
\subsection{OTFS Transmitter and Frame Structure}
At the transmitter, information bits are mapped to symbols as $\{x_{d}[k,l]:k=0,1,\cdots,N-1,l=0,1,\cdots,M-1\}$ on the delay-Doppler grid. $x_{d}[k,l]$ are then converted to time-frequency symbols $X_{d}[n,m]$ employing the inverse symplectic finite Fourier transform (ISFFT). Let $\tau_{\text{max}}$ and $\nu_{\text{max}}$ denote the maximum delay and Doppler spread corresponding to $f_{c}$, where we have $l_{\text{max}}=\lceil\tau_{\text{max}}M\Delta f\rceil$ and $k_{\text{max}}=\lceil\nu_{\text{max}}NT\rceil$. \par
\begin{figure}
	\centering
	\includegraphics[width=0.9\linewidth]{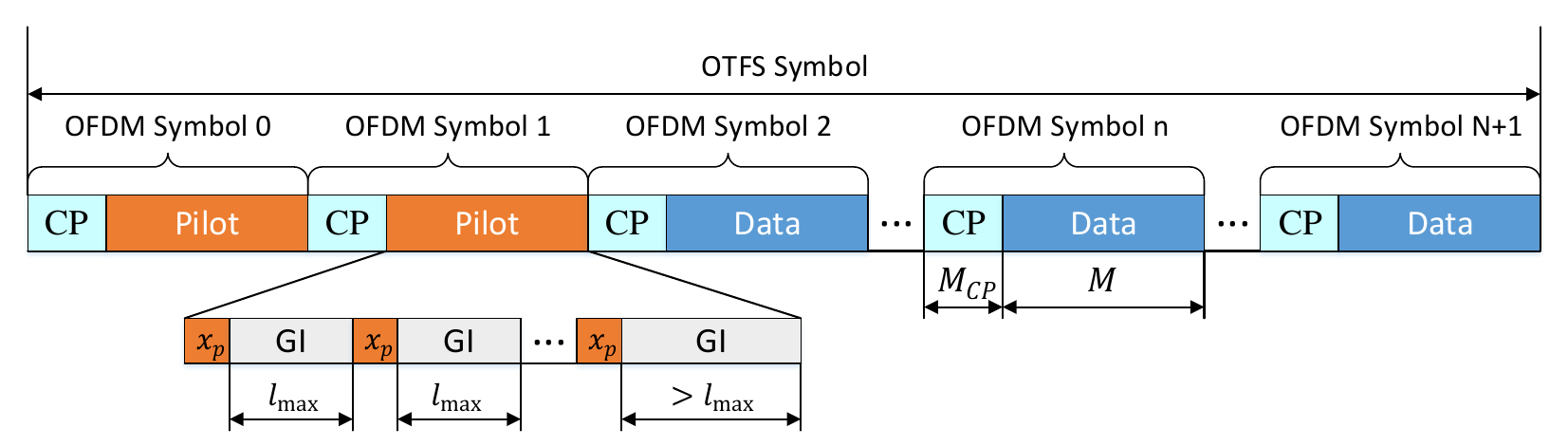}
	\caption{OTFS frame structure in the time domain.}
	\label{Fig_frame}	
\end{figure} 
CP-OFDM is adopted as the pulse-shape to create the continuous baseband waveform. $T_{CP}=\frac{M_{CP}}{M}T$ and $T_{u}=\frac{M+M_{CP}}{M}T$ represent the time duration of a CP and an entire OFDM symbol, where we have $T=\frac{1}{\Delta f}$ to guarantee the orthogonality. OTFS frame structure in the time domain is depicted in Fig. \ref{Fig_frame}, where two prefix OFDM symbols are allocated for channel estimation. Let $X[n,m]$ denote time-frequency symbols including the pilot and data which is loaded as $X[n+2,m]=X_{d}[n,m]$. The transmitted waveform in the time domain can be derived as
\begin{equation}
	\small
	s(t)=\frac{1}{\sqrt{M}}\sum_{n=0}^{N+1}\sum_{m=0}^{M-1}X[n,m]g(t-nT_{u}+T_{CP})e^{j2\pi m\Delta f(t-nT_{u})},
	\label{transmitted_waveform}
\end{equation}
where $g(t)$ is the unit rectangular pulse of duration $T_{u}$.\par 
$s(t)$ can be sampled as $S[n,l]=s(nT_{u}+\frac{l}{M}T)$ like \cite{OTFS_CE_EMVB}, where we have
\begin{equation}
	S[n,l]=\frac{1}{\sqrt{M}}\sum_{m=0}^{M-1}X[n,m]e^{j2\pi\frac{ml}{M}}.
	\label{Tx_time_discrete}
\end{equation}
Let $x_{p}$ and $M_{p}=\lfloor\frac{M}{l_{\text{max}}+1} \rfloor$ represent the pilot value and the number of pilots within an OFDM symbol, where we have
\begin{equation}
	\small
	S[n,l]=
	\begin{cases}
		x_{p},&l=q(l_{\text{max}}+1),q=0,1,\cdots,M_{p}-1\\
		0,&\text{elsewhere}
	\end{cases}
\label{Tx_time_pilot}
\end{equation}
for $n=0,1$. When it comes to OFDM symbols where data is loaded, we can derive 
\begin{equation}
	\small
	S[n+2,l]=\frac{1}{\sqrt{N}}\sum_{k=0}^{N-1}x_{d}[k,l]e^{j2\pi\frac{nk}{N}}, n=0,\cdots,N-1.
	\label{Tx_time_data}
\end{equation} 
%to build the relationship between the time-delay domain and the delay-Doppler domain. 
\subsection{Multipath LTV Channel Model with DSE}
The passband waveform $\Re\{s(t)e^{j2\pi f_{c}t}\}$ travels from the transmitter to the receiver via $N_{P}$ incident paths. The received passband signal is \cite{channel_time_orig,OTFS_DSE_mine}
\begin{equation}
	\small
	\widetilde{r}(t)=\Re \bigg\{\sum_{i=1}^{N_{P}}\widetilde{\beta_{i}}s(t-(\tau_{i}-\frac{v_{i}}{c}t))e^{j2\pi f_{c}(t-(\tau_{i}-\frac{v_{i}}{c}t))}\bigg\},
	\label{Rx_passband_signal}
\end{equation} 
where $v_{i}$ and $c$ represent the velocity with which the $i^{th}$ path length decreases and the speed of light, respectively. $\widetilde{\beta_{i}}$, $\tau_{i}$ denote the attenuation and propagation delay of the $i^{th}$ path, respectively. Let $\nu_{i}=\frac{v_{i}}{c}f_{c}$ denote the Doppler shift at $f_{c}$ and remove the carrier $e^{j2\pi f_{c}t}$, the baseband received signal can be written as
\begin{equation}
	\small
	r(t)=\sum_{i=1}^{N_{P}}\widetilde{\beta_{i}}e^{-j2\pi \tau_{i}f_{c}}e^{j2\pi\nu_{i}t}s(t-(\tau_{i}-\frac{\nu_{i}}{f_{c}}t)).
	\label{Rx_baseband_signal_DSE}
\end{equation}\par
For OFDM systems, the frame duration is $T$ while the sampling period is $\frac{T}{M}$, which deduces $\frac{\nu_{i}}{f_{c}}T\ll\frac{T}{M}\leqslant\tau_{i}$ even though in high-mobility scenarios with large number of subcarriers, e.g., $v=1000$km/h and $M=1024$. Therefore, $\tau_{i}-\frac{\nu_{i}}{f_{c}}t\approx \tau_{i}$ holds true, which derives the channel model employed in \cite{OTFS_cross_sigmodel}. However, in OTFS systems, the frame duration becomes longer than $NT$. Meanwhile, $N$ is usually large to enjoy full-time diversity, e.g., $N=128$ \cite{OTFS_cross_sigmodel,OTFS_CE_MP}. Therefore, $\frac{\frac{\nu_{i}}{f_{c}}NT}{T/M}$ is about $0.13$ for $M=1024$, $N=128$ and $v=1000$km/h, which is non-negligible. \par 
On the other hand, the time-variant frequency response $H(t,f)$ is also widely employed to characterize the LTV channel. The received signal can be derived \cite{OTFS_DSE_mine} as $r(t)=\int H(t,f)S(f)e^{j2\pi tf}df,
\label{LTV_Htf_general}$, where $S(f)$ is the Fourier transform of $s(t)$. Combining with \eqref{Rx_baseband_signal_DSE}, $H(t,f)$ can be derived as
\begin{equation}
	\small
	H(t,f)=\sum_{i=1}^{N_{P}}
	\beta_{i}
	e^{j2\pi\frac{\nu_{i}}{f_{c}}(f_{c}+f)t} e^{-j2\pi f\tau_{i}},
	\label{LTV_Htf_DSE}
\end{equation} 
where we have $\beta_{i}=\widetilde{\beta_{i}}e^{-j2\pi f_{c}\tau_{i}}$ to simplify the notation. \eqref{LTV_Htf_DSE} indicates that the Doppler shift brought by the high-mobility in multipath LTV channel is $\frac{\nu_{i}}{f_{c}}(f_{c}+f)$, which is frequency-dependent and referred to as DSE. In OTFS systems, the subcarrier-dependent phase offset brought by DSE will be accumulated within one OTFS symbol as $e^{j2\pi\frac{\nu_{i}}{f_{c}}ft}$. Taking the typical value as $M=1024$, $N=128$ and $v=1000$km/h, DSE leads to a maximum offset about $e^{j2\pi\frac{\nu_{i}}{f_{c}}\times M\Delta f \times NT}\approx e^{j0.26\pi}$, which plays a significant role in OTFS systems. If ignored directly, striking performance degradation occurs \cite{OTFS_DSE_mine}.\par 
Moreover, DSE leads to time-frequency coupling and destroys the sparsity in delay-Doppler channel response, which is disastrous to most of existing designs since the sparsity of the delay-Doppler channel is usually treated as the basic assumption. According to \cite{OTFS_DSE_mine}, the delay-Doppler response $h(\tau,\nu)$ of multipath LTV channel for $\nu_{i}\neq 0$ is $h(\tau,\nu) = \sum_{i=1}^{N_{P}}\beta_{i}|p_{i}|e^{j2\pi p_{i}(\tau-\tau_{i})(\nu-\nu_{i})}$, where we have $p_{i}=\frac{f_{c}}{\nu_{i}}$ for ease of illustration. It is significantly different from the channel model $h(\tau,\nu)=\sum_{i=1}^{N_{P}}\beta_{i}\delta(\tau-\tau_{i})\delta(\nu-\nu_{i})$ in \cite{OTFS_cross_sigmodel,OTFS_cross_sigmodel2}, which inspires us to reconsider the input-output analysis and receiver design in prior work for practical OTFS systems with DSE. 
\section{OTFS Input-Output Analysis with DSE}
At the receiver, $r(t)$ is sampled as $R[n,l]=r(nT_{u}+\frac{l}{M}T)$ like $S[n,l]$. In this section, we investigate the CP length to ensure no ISI between OFDM symbols and derive the input-output relation with DSE. Meanwhile, notations $k_{i}=\nu_{i}NT$ and $l_{i}=\tau_{i}M\Delta f$ are employed, where $l_{i}\geqslant 1$ is required to exploit the delay resolution. Similar to \cite{OTFS_DSE_mine}, OTFS system with $(N+2)M<10^{6}<|p_{i}|$ is adopted in this paper, which is easily compatible with the existing wireless communication network \cite{OTFS_cross_sigmodel}, e.g., $M=1024$, $N=128$ with the relative velocity $v\leqslant1000$km/h.
\subsection{Requirement of CP Length}   
In order to simplify the equalization by employing the block diagonal property of channel matrix like \cite{OTFS_LMMSEdata1,OTFS_CPOFDM_sigmodel_access}, CP length is required to be adequate to confirm no ISI between OFDM symbols, which is illustrated in detail in this subsection.\par 
Combining \eqref{Rx_baseband_signal_DSE} with \eqref{transmitted_waveform} at the sampled point, the requirement of CP length is
\begin{equation}
	\small
	(1+\frac{1}{p_{i}})(nT_{u}+\frac{l}{M}T)-\frac{l_{i}}{M}T-nT_{u}+T_{CP}\in (0,T_{u}), \forall n,l,i,
	\label{CPana_ini}
\end{equation}
which is obtained from the duration of $g(t)$. Let us begin with the analysis of the left part, where \eqref{CPana_ini} can be simplified as
\begin{equation}
	\small
	T_{CP}>\frac{l_{i}}{M}T+\frac{(N+1)T_{u}+\frac{l}{M}T}{p_{i}}-\frac{l}{M}T, \forall l,i.
	\label{CPana_0_1}
\end{equation}
By employing $|p_{i}|>(N+2)M$, we have
\begin{equation}
	\small
	\begin{aligned}
		&\frac{l_{i}}{M}T+\frac{(N+1)T_{u}+\frac{l}{M}T}{|p_{i}|}-\frac{l}{M}T\\
		&<\frac{l_{\text{max}}}{M}T+\frac{(N+1)(T+T_{CP})+T}{(N+2)M}\\
		&\overset{(a)}{<}\frac{l_{\text{max}}+\frac{2N+3}{N+2}}{M}T<\frac{l_{\text{max}}+2}{M}T,
	\end{aligned}
\label{CPana_0_process}
\end{equation}
where (a) is attained by assuming $T_{CP}<T$.
\eqref{CPana_0_process} means $M_{CP}\geqslant l_{\text{max}}+2$ is sufficient to ensure \eqref{CPana_0_1}.\par 
On the other hand, we can derive
\begin{equation}
	\small
	\begin{aligned}
		&(1+\frac{1}{p_{i}})(nT_{u}+\frac{l}{M}T)-\frac{l_{i}}{M}T-nT_{u}+T_{CP}-T_{u}\\
		&=\frac{l}{M}T+T_{CP}-T_{u}-\frac{l_{i}}{M}T+\frac{nT_{u}+\frac{l}{M}T}{p_{i}}\\
		&<-\frac{1}{M}T-\frac{l_{i}}{M}T+\frac{(N+2)T+(N+1)T_{CP}}{(N+2)M}\\
		&\overset{(a)}{<}\frac{T}{M}(-1-1+\frac{2N+3}{N+2})<0,
	\end{aligned}
\label{CPana_Tu_process}
\end{equation}
where (a) is obtained by employing $T_{CP}<T$ and $l_{i}\geqslant 1$. \eqref{CPana_Tu_process} indicates that no additional conditions are demanded to achieve \eqref{CPana_ini}. Therefore, if the length of CP satisfies
\begin{equation}
	\small
	l_{\text{max}}+2\leqslant M_{CP}<M,
	\label{CP_condition}
\end{equation}
there will be no ISI between OFDM symbols, which simplifies the receiver design significantly. \par 
\subsection{OTFS Input-Output Analysis with DSE}
In this subsection, the input-output analysis is offered, where the CP length satisfies \eqref{CP_condition}. The relation between $R[n,l]$ and $S[n,l]$ is provided by the following theorem.\par 
\begin{theorem}
	\rm
	\label{th1_h_S2R}
   The input-output relation of CP-OFDM-based OTFS systems can be represented as
	\begin{equation}
		%\small
		R[n,l]=\sum_{i=1}^{N_{P}}\sum_{l^{\prime}=0}^{M-1}h^{i}_{n}[l,l^{\prime}]S[n,l^{\prime}],
		\label{S2R_sum}
	\end{equation}
where $h^{i}_{n}[l,l^{\prime}]$ is formulated in \eqref{h_time_S2R}. 
	\begin{IEEEproof}
		The proof is provided in Appendix \ref{th1_proof_apendix}.
	\end{IEEEproof}
\end{theorem}
\begin{figure*}
	\begin{equation}
		\small
		h^{i}_{n}[l,l^{\prime}]=\beta_{i}e^{j2\pi\left(\frac{n(M+M_{CP})+l}{M}\frac{k_{i}}{N}(1+\frac{(M-1)\Delta f}{2f_{c}})\right)}e^{j\pi\frac{M-1}{M}(l-l^{\prime}-l_{i})}\frac{\sin{\pi(l-l^{\prime}-l_{i}+\frac{n(M+M_{CP})+l}{p_{i}})}}{M\sin{\frac{\pi}{M}(l-l^{\prime}-l_{i}+\frac{n(M+M_{CP})+l}{p_{i}})}}
		\label{h_time_S2R}
	\end{equation}
\hrulefill
\end{figure*}
Compared with prior analysis in \cite{OTFS_CE_EMVB} where $p_{i}$ is treated as $\infty$ to ignore DSE, two modifications occur in \eqref{h_time_S2R}: 
\begin{itemize}
	\item [1)] Delay spread extension: If $p_{i}$ in the phase of sinc function is ignored, $h_{n}^{i}[l,l^{\prime}]=0$ holds true for $\forall l\neq (l^{\prime}+l_{i})_{M}$ when integer delay $l_{i}$ is involved. However, extra phase $\frac{n(M+M_{CP})+l}{p_{i}}$ destroys this property and brings more delay spread. Though the location of the most powerful channel coefficient seldom changes due to DSE, significant delay spread extension aggravates the interference within an OFDM symbol especially when $n$ is large.
	\item [2)] Extra phase shift: DSE introduces an extra phase shift besides traditional Doppler shift, which is represented as
	\begin{equation}
		\small
		e^{j2\pi\frac{n(M+M_{CP})+l}{M}\frac{k_{i}}{N}\frac{(M-1)\Delta f}{2f_{c}}}=e^{j\pi\frac{M-1}{M}\frac{n(M+M_{CP})+l}{p_{i}}}.
	\end{equation}
Taking the typical value as $N=128$, $M=1024$, $M_{CP}=24$ and the relative velocity as $v=1000$km/h, DSE leads to a maximum phase offset of about $0.13\pi$.
\end{itemize}\par 
From the analysis above, the significance of DSE is determined by $\frac{n(M+M_{CP})+l}{p_{i}}$, which can be approximately measured by $\frac{(N+2)M}{p_{i}}$. It is different from prior declaration \cite{DSE_air} where DSE is caused only by large bandwidth. Since the Doppler squint accumulates through the time, the significance is determined by the size of the entire time-frequency block rather than the bandwidth. e.g., if the bandwidth is fixed, more subcarriers mean smaller subcarrier spacing, which increases $T$ and consequently amplifies the significance of DSE. As a result, the impact of DSE is dependent on the ratio between the time-frequency resource block size $(N+2)T\times M\Delta f=(N+2)M$ and the mobility parameter $p_{i}=\frac{f_{c}}{\nu_{i}}=\frac{c}{v_{i}}$.\par 
Meanwhile, DSE disappears when $\nu_{i}=0$, whose input-output formulation can be treated as the limit for \eqref{h_time_S2R} when $p_{i}\rightarrow\infty$. It is worth pointing out that the delay-Doppler input-output relationship can also be attained by Theorem \ref{th1_h_S2R}. However, it is not depicted since no closed-form representation can be achieved and the derivation cannot benefit the receiver design in this paper. It is also valuable to focus on the delay-Doppler response and provide the corresponding receiver schemes, where similar characteristics appear like \cite{OTFS_DSE_mine}.
\section{Low-Complexity Receiver Design with DSE}
In this section, the low-complexity OTFS receiver design including the parameter extraction-based channel recovery and the LMMSE-based data detection is illustrated in detail. Similar to \cite{OTFS_CE_MP,OTFS_LMMSEdata1,OTFS_CE_thres1,OTFS_CE_thres3,OTFS_CE_EMVB,OTFS_cross_sigmodel,OTFS_DSE_mine}, the system bandwidth is assumed to be sufficient to eliminate the fractional delay, where $l_{i}$ are all integers while $k_{i}$ are not necessarily integers.
\subsection{Channel Estimation Scheme}
At first, $\left|\frac{n(M+M_{CP})+l}{p_{i}}\right|<\frac{3M}{(N+2)M}=\frac{3}{N+2}\ll 1$ can be derived for OFDM symbols employed for channel estimation since $n\leqslant1$, $l\leqslant M-1$ and $|p_{i}|>(N+2)M$ hold true. It indicates that the phase of sinc function in \eqref{h_time_S2R} can be approximated by $l-l^{\prime}-l_{i}$. Therefore, if there is not a path with delay $l_{i}$, we can deduce
\begin{equation}
	\small
	R[n,q(l_{\text{max}}+1)+l_{i}]\approx 0,n=0,1,q=0,1,\cdots,M_{p}-1.
	\label{R_CE_nopath}
\end{equation}
Otherwise, the received symbols can be represented as
\begin{equation}
	%\small
	\begin{aligned}
		&R[n,q(l_{\text{max}}+1)+l_{i}]\approx x_{p}\beta_{i}\\
		&\times e^{j2\pi\left(\frac{n(M+M_{CP})+q(l_{\text{max}}+1)+l_{i}}{M}\frac{k_{i}}{N}(1+\frac{(M-1)\Delta f}{2f_{c}})\right)}
	\end{aligned}
\label{R_CE_pathli}
\end{equation}
for $n=0,1$ if there is a path with $l_{i}$, $k_{i}$ and $\beta_{i}$, which inspires us to judge the existence of path with $l_{i}$ by employing the threshold-based detection. We employ the vectorized notation $\mathbf{r}_{n}^{l_{i}}\in \mathbb{C}^{M_{p}\times 1}$ as $\mathbf{r}_{n}^{l_{i}}(q)=R[n,q(l_{\text{max}}+1)+l_{i}]$ for $n=0,1$. $|\mathbf{r}_{n}^{l_{i}}(q)|>\Gamma$ for $\forall n=0,1,q=0,1,\cdots,M_{p}-1$ indicates there is a path with delay $l_{i}$.\par 
After detecting the path with $l_{i}$, $k_{i}$ can be extracted by phase differences. Let $\boldsymbol{\theta}\in\mathbb{R}^{M_{p}\times 1}$ denote the phase difference array which is computed as
\begin{equation}
	\small
	\boldsymbol{\theta}(q)=\angle{\frac{\mathbf{r}_{1}^{l_{i}}(q)}{\mathbf{r}_{0}^{l_{i}}(q)}}.
	\label{phase_difference_array}
\end{equation}
If ignoring the noise, $\overline{\boldsymbol{\theta}}=2\pi\frac{M+M_{CP}}{M}\frac{k_{i}}{N}(1+\frac{(M-1)\Delta{f}}{2f_{c}})$ holds true, which motivates the approximate extraction of $\hat{k}_{i}$ as
\begin{equation}
	\small
	\hat{k}_{i}=\frac{MN}{2\pi(M+M_{CP})(1+\frac{(M-1)\Delta{f}}{2f_{c}})}\overline{\boldsymbol{\theta}}.
	\label{estimation_ki}
\end{equation}\par
Finally, the least-square (LS) solution is employed to estimate $\beta_{i}$, where the base vector $\boldsymbol{\psi}_{n}^{l_{i}}\in\mathbb{C}^{M_{p}\times 1}$ are
\begin{equation}
	\boldsymbol{\psi}_{n}^{l_{i}}(q)=x_{p}e^{j2\pi\frac{n(M+M_{CP})+q(l_{\text{max}}+1)+l_{i}}{M}\frac{\hat{k}_{i}}{N}(1+\frac{(M-1)\Delta f}{2f_{c}})}
\end{equation}
for $n=0,1$. $\hat{\beta}_{i}$ can be easily attained by
\begin{equation}
	\small
	\hat{\beta}_{i}=\left(\boldsymbol{\psi}_{i}^{H}\boldsymbol{\psi}_{i}\right)^{-1}\boldsymbol{\psi}_{i}^{H}\mathbf{r}_{i},
	\label{estimation_betai}
\end{equation}
where $\boldsymbol{\psi}_{i}=\left[
\begin{matrix}
	\boldsymbol{\psi}_{0}^{l_{i}}\\
	\boldsymbol{\psi}_{1}^{l_{i}}
\end{matrix}
\right],
\mathbf{r}_{i}=\left[
\begin{matrix}
	\mathbf{r}_{0}^{l_{i}}\\
	\mathbf{r}_{1}^{l_{i}}
\end{matrix}
\right]$. After the parameter extraction of $l_{i}$, $\hat{k}_{i}$ and $\hat{\beta}_{i}$, CSI can be recovered by Theorem \ref{th1_h_S2R}. 
\subsection{LMMSE-based Equalization Scheme}
\begin{algorithm}[t]
	%\small
	\renewcommand{\algorithmicrequire}{\textbf{Input:}}
	\renewcommand{\algorithmicensure}{\textbf{Output:}}
	\caption{Proposed OTFS Receiver design with DSE}
	\label{alg:1}
	\begin{algorithmic}[1]
		\REQUIRE
		$R[n,l]$, $l_{\text{max}}$, $x_{p}$, $\Gamma$, $\sigma_{n}^{2}$ and $\sigma_{s}^{2}$.
		\STATE
		\textbf{Channel Estimation:}
		\STATE
		\textbf{Initialize} $\hat{\mathbf{l}}$, $\hat{\mathbf{k}}$ and $\hat{\boldsymbol{\beta}}$ as empty parameter vectors;
		\FOR{$l_{i}=1:l_{\text{max}}$}
		\STATE 
		Generate $\mathbf{r}_{0}^{l_{i}}$ and $\mathbf{r}_{1}^{l_{i}}$ by $\mathbf{r}_{n}^{l_{i}}(q)=R[n,q(l_{\text{max}}+1)+l_{i}]$;
		\STATE
		\% Detect the path with delay $l_{i}$
		\IF
		{$\mathop{\min}_{n=0,1,q=0,1,\cdots,M_{p}-1}\left|\mathbf{r}_{n}^{l_{i}}(q)\right|>\Gamma$}
		\STATE
		Successfully detect the path with delay $l_{i}$;
		\STATE
		Extract the normalized Doppler $\hat{k}_{i}$ by \eqref{phase_difference_array} and \eqref{estimation_ki};
		\STATE
		Estimate $\hat{\beta}_{i}$ by \eqref{estimation_betai};
		\STATE
		Add $l_{i}$, $\hat{k}_{i}$ and $\hat{\beta}_{i}$ to $\hat{\mathbf{l}}$, $\hat{\mathbf{k}}$ and $\hat{\boldsymbol{\beta}}$, respectively;
		\ENDIF
		\ENDFOR
		\STATE
		Employ $\hat{\mathbf{l}}$, $\hat{\mathbf{k}}$ and $\hat{\boldsymbol{\beta}}$ to recover the channel matrix $\hat{\mathbf{H}}_{n}$; 
		\STATE
		\textbf{Data detection:}
		\FOR{$n=2:N+1$}
		\STATE
		Carry out the equalization as \eqref{estimation_sn} to recover $\hat{\mathbf{s}}_{n}$;
		\ENDFOR
		\STATE
		Attain $\hat{x}_{d}[k,l]$ by utilizing \eqref{estimation_xd};
		\STATE
		Obtain information bits from $\hat{x}_{d}[k,l]$;
		\STATE
		\textbf{Return} information bits and possibly $\hat{\mathbf{H}}_{n}$;
	\end{algorithmic}		
\end{algorithm}
Since no ISI between OFDM symbols is introduced, it is much easier to carry out the equalization for each OFDM symbol. The vectorized input-output relationship is employed as $\mathbf{r}_{n}=\mathbf{H}_{n}\mathbf{s}_{n}+\mathbf{w}_{n}$, where we have $\mathbf{r}_{n}(l)=R[n,l]$, $\mathbf{s}_{n}(l)=S[n,l]$ and $\mathbf{H}_{n}(l,l^{\prime})=\sum_{i=1}^{N_{P}}h_{n}^{i}[l,l^{\prime}]$. $\mathbf{w}_{n}\sim\mathcal{CN}(\mathbf{0},\sigma_{n}^{2}\mathbf{I}_{M})$ denotes the additive white Gaussian noise.\par 
The LMMSE-based equalizer is then implemented as
\begin{equation}
	\small
	\hat{\mathbf{s}}_{n}=\left(\hat{\mathbf{H}}_{n}^{H}\hat{\mathbf{H}}_{n}+\frac{\sigma_{n}^{2}}{\sigma_{s}^{2}}\mathbf{I}\right)^{-1}\hat{\mathbf{H}}_{n}^{H}\mathbf{r}_{n},
	\label{estimation_sn}
\end{equation}
where $\mathbf{H}_{n}$ are recovered as $\hat{\mathbf{H}}_{n}$ based on the estimated parameters. $\sigma_{s}^{2}$ represents the average power of $x_{d}[k,l]$, which is deduced by employing the property of IFFT in \eqref{Tx_time_data}. Delay-Doppler symbols can therefore be recovered by
\begin{equation}
	%\small
	\hat{x}_{d}[k,l]=\frac{1}{\sqrt{N}}\sum_{n=0}^{N-1}\hat{S}[n+2,l]e^{-j2\pi\frac{nk}{N}},
	\label{estimation_xd}
\end{equation} 
which can be directly employed to recover information bits.
\subsection{Low-Complexity Receiver Design}
The low-complexity OTFS receiver design with DSE is briefly drawn in this subsection. As illustrated in \textbf{Algorithm \ref{alg:1}}, the parameter estimation-based CSI recovery is carried out first, which can be divided into three steps as delay detection, Doppler extraction and gain estimation. The estimated parameters are then employed to recover the channel matrices $\hat{\mathbf{H}}_{n}$, which are utilized in the LMMSE-based equalizer. At last, delay-Doppler symbols $x_{d}[k,l]$ can be estimated and information bits can be obtained, which finishes the design of OTFS receiver considering DSE.\par 
For each iteration of channel estimation, the computational complexity is bounded as $\mathcal{O}\left(M_{p}\right)$, which causes a total load of $\mathcal{O}\left(M\right)$ for the parameter extraction. When it comes to the data detection, complexity of $\mathcal{O}\left(NM^{3}+M\log_{2}{N}\right)\approx\mathcal{O}\left(NM^{3}\right)$ is required due to the inversion operation in each iteration. Taking the excellent structure for parallel processing in \textbf{Algorithm \ref{alg:1}}, the total complexity can be bounded as $\mathcal{O}\left(M^{3}+M_{p}+\log_{2}{N}\right)\approx\mathcal{O}\left(M^{3}\right)$, which is quite lower than the delay-Doppler implementation of LMMSE-based receiver with $\mathcal{O}\left(N^{3}M^{3}\right)$. It is worth pointing out that though plenty of works have been devoted to the LMMSE-based receiver \cite{OTFS_LMMSEdata2,OTFS_LMMSEdata1} with lower complexity, the negligence of DSE leads to the failure of these schemes, which inspires future work to further explore the channel structure of DSE and develop more efficient receiver designs.   
\section{Simulation Results}
\begin{table}
	\caption{Simulation Parameters}
	\centering
	\label{simulation_para_table}
	\renewcommand\arraystretch{1.2}
	\begin{tabular}{p{15em}p{10em}}
		\hline
		Parameter &
		Typical value\\
		\hline
		Carrier frequency ($f_{c}$)& $4$GHz\\
		Subcarrier spacing ($\Delta f$)& $30$kHz\\
		Number of subcarriers ($M$)& 1024\\
		Number of OFDM symbols ($N$)& 128\\
		UE speed (km/h)& 1000\\
		Number of paths ($N_{P}$)& 4\\
		Maximum delay grid ($l_{\text{max}}$)& 20\\
		Length of CP ($M_{CP}$)& 24\\
		Modulation alphabet & QPSK\\
		\hline
	\end{tabular}
\end{table}
The significance of DSE and the performance of the proposed receiver scheme are evaluated in this section by presenting simulation results. Jakes' formula is utilized to generate the Doppler shift of each path as $\nu_{i}=\nu_{\text{max}}\cos{\theta_{i}}$ where $\theta_{i}$ is uniformly distributed over $\left[-\pi,\pi\right]$. Complex coefficients $\beta_{i}$ are generated as $\beta_{i}\sim\mathcal{CN}\left(0,1/N_{P}\right)$. The normalized mean squared error (NMSE) of the channel matrix is defined as
\begin{equation}
	%\small
	\text{NMSE}=\mathbb{E}\frac{||\hat{\mathbf{H}}_{N+1}-\mathbf{H}_{N+1}||_{F}^{2}}{||\mathbf{H}_{N+1}||_{F}^{2}},
	\label{NMSE}
\end{equation}
which is employed to evaluate the error of both the model and the estimation. We set $\Gamma=3\sigma_{n}$ and limit the absolute value of $\hat{k}_{i}$ by $k_{\text{max}}$ to implement \textbf{Algorithm \ref{alg:1}}. The system signal-to-noise ratio is defined as SNR$=\frac{\sigma_{s}^{2}}{\sigma_{n}^{2}}$ while the power of pilot symbols is 30dB higher than $\sigma_{s}^{2}$.\par 
\begin{figure}
	\centering
	\includegraphics[width=0.75\linewidth]{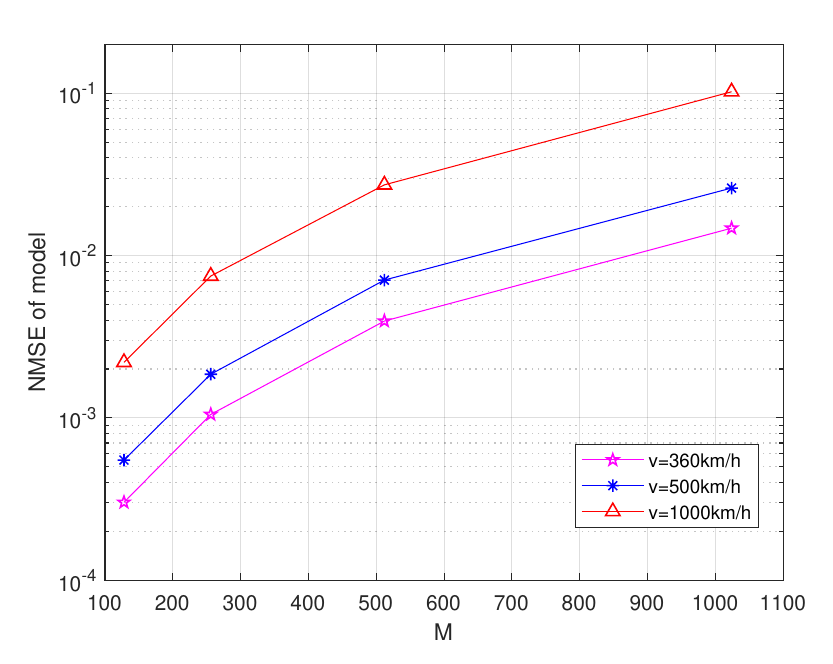}
	\caption{NMSE against $M$ with perfect knowledge of channel parameters.}
	\label{Fig_NMSE_model}	
\end{figure}
The significance of DSE is first verified by plotting NMSE of model in Fig. \ref{Fig_NMSE_model}, where $\hat{\mathbf{H}}_{N+1}$ in \eqref{NMSE} denotes the channel matrix ignoring DSE with perfect knowledge of parameters. It is clear that the impact of DSE increases with the velocity and $M$ increasing. NMSE of about 10\% occurs when $v=1000$km/h and $M=1024$, which deserves serious consideration to ensure the reliability in practical OTFS systems.\par 
\begin{figure}
	\centering
	\includegraphics[width=0.75\linewidth]{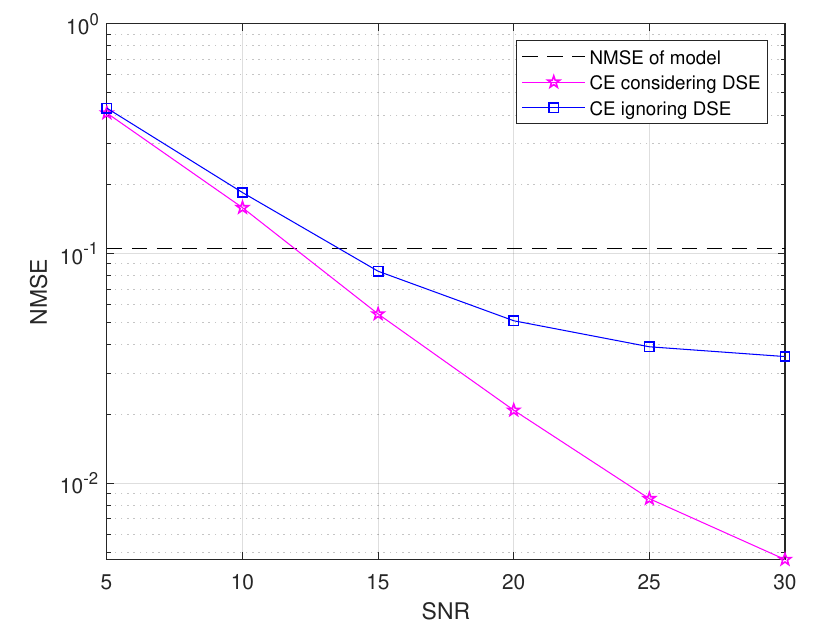}
	\caption{NMSE performance against SNR with parameters set as Table \ref{simulation_para_table}.}
	\label{Fig_NMSE_CE}	
\end{figure}
\begin{figure}
	\centering
	\includegraphics[width=0.75\linewidth]{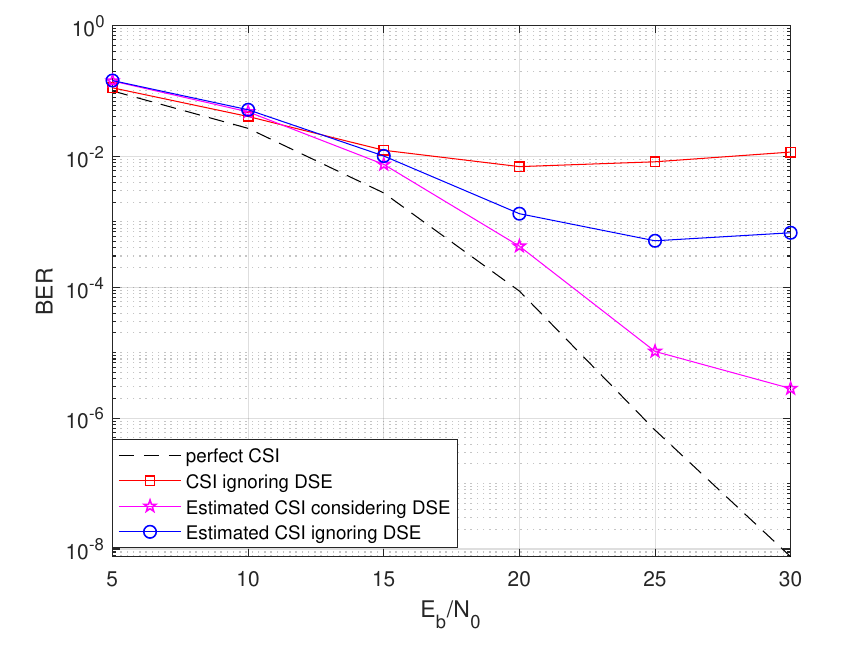}
	\caption{BER performance against SNR with parameters set as Table \ref{simulation_para_table}.}
	\label{Fig_ber}	
\end{figure}
Fig. \ref{Fig_NMSE_CE} demonstrates the performance of channel estimation by presenting the NMSE comparison. When SNR$=15$dB, NMSE of the estimated CSI considering DSE approaches half of NMSE of model. NMSE of estimated CSI considering DSE is less than $1\%$ when SNR$\geqslant25$dB. However, NMSE is more than $3\%$ when SNR$=30$dB if neglecting DSE, which confirms the importance of considering DSE again.\par  
\begin{figure*}[t]
	\begin{equation}
		\small
		\begin{aligned}
			R_{i}[n,l]&=r_{i}(nT_{u}+\frac{l}{M}T)=\beta_{i}e^{j2\pi\nu_{i}(nT_{u}+\frac{l}{M}T)}s(nT_{u}+\frac{l}{M}T-\tau_{i}+\frac{nT_{u}+\frac{l}{M}T}{p_{i}})\\
			&=\frac{\beta_{i}}{\sqrt{M}}e^{j2\pi\frac{k_{i}\left( n(M+M_{CP})+l\right)}{NM}}\sum_{m=0}^{M-1}X[n,m]e^{j2\pi m\Delta{f}\left(\frac{l-l_{i}}{M}T+\frac{nT_{u}+\frac{l}{M}T}{p_{i}}\right)}\\
			&=\frac{\beta_{i}}{M}e^{j2\pi\frac{k_{i}\left( n(M+M_{CP})+l\right)}{NM}}\sum_{m=0}^{M-1}\sum_{l^{\prime}=0}^{M-1}S[n,l^{\prime}]e^{-j2\pi\frac{ml^{\prime}}{M}}e^{j2\pi\frac{m}{M}\left(l-l_{i}+\frac{n(M+M_{CP})+l}{p_{i}}\right)}\\
			&=\frac{\beta_{i}}{M}e^{j2\pi\frac{k_{i}\left( n(M+M_{CP})+l\right)}{NM}}\sum_{l^{\prime}=0}^{M-1}S[n,l^{\prime}]\sum_{m=0}^{M-1}e^{j2\pi\frac{m}{M}\left(l-l^{\prime}-l_{i}+\frac{n(M+M_{CP})+l}{p_{i}}\right)}\\
			&=\frac{\beta_{i}}{M}e^{j2\pi\frac{k_{i}\left( n(M+M_{CP})+l\right)}{NM}}\sum_{l^{\prime}=0}^{M-1}S[n,l^{\prime}]e^{j\pi\frac{M-1}{M}\left(l-l^{\prime}-l_{i}+\frac{n(M+M_{CP})+l}{p_{i}}\right)}\frac{\sin{\pi\left(l-l^{\prime}-l_{i}+\frac{n(M+M_{CP})+l}{p_{i}}\right)}}{\sin{\frac{\pi}{M}\left(l-l^{\prime}-l_{i}+\frac{n(M+M_{CP})+l}{p_{i}}\right)}}\\
			&=\sum_{l^{\prime}=0}^{M-1}S[n,l^{\prime}]\beta_{i}e^{j2\pi\frac{n(M+M_{CP})+l}{M}\left(\frac{k_{i}}{N}+\frac{M-1}{2p_{i}}\right)}e^{j\pi\frac{M-1}{M}\left(l-l^{\prime}-l_{i}\right)}\frac{\sin{\pi\left(l-l^{\prime}-l_{i}+\frac{n(M+M_{CP})+l}{p_{i}}\right)}}{M\sin{\frac{\pi}{M}\left(l-l^{\prime}-l_{i}+\frac{n(M+M_{CP})+l}{p_{i}}\right)}}\\
			&=\sum_{l^{\prime}=0}^{M-1}\underbrace{\beta_{i}e^{j2\pi\frac{n(M+M_{CP})+l}{M}\frac{k_{i}}{N}(1+\frac{(M-1)\Delta f}{2f_{c}})}e^{j\pi\frac{M-1}{M}(l-l^{\prime}-l_{i})}\frac{\sin{\pi(l-l^{\prime}-l_{i}+\frac{n(M+M_{CP})+l}{p_{i}})}}{M\sin{\frac{\pi}{M}(l-l^{\prime}-l_{i}+\frac{n(M+M_{CP})+l}{p_{i}})}}}_{h_{n}^{i}[l,l^{\prime}]}  S[n,l^{\prime}]
		\end{aligned}
		\label{S2R_derive_process}
	\end{equation}
	\hrulefill
\end{figure*}

In Fig. \ref{Fig_ber}, BER comparison is displayed to show the performance superiority of the receiver considering DSE. When SNR$\geqslant20$dB, BER floor of about $10^{-2}$ appears because of the negligence of DSE even perfect knowledge of parameters is provided. When estimated CSI ignoring DSE is offered, BER floor can be reduced to about $5\times10^{-4}$, which corresponds to the NMSE performance in Fig. \ref{Fig_NMSE_CE}. However, when the estimated CSI considering DSE is employed, BER floor can be diminished with BER of less than $5\times10^{-4}$ when SNR$\geqslant20$dB, which approaches the performance with perfect CSI.
\section{Conclusion}
In this paper, CP-OFDM-based OTFS systems with DSE are investigated for the first time. Based on the analysis of CP length and the derivation of input-output relationship, a low-complexity receiver design is offered to realize the channel estimation and data detection in practical OTFS systems with DSE. Simulation results demonstrate the significance of DSE and the appreciable performance of the proposed receiver considering DSE. For future research, it is valuable to consider the optimization of the pilot and data detection where higher transmission efficiency and lower complexity can be obtained.  
\appendices
\section{Proof of Theorem \ref{th1_h_S2R}}
\label{th1_proof_apendix}
Since \eqref{CP_condition} is satisfied, no ISI occurs between OFDM symbols. Similar to \cite{OTFS_DSE_mine,OTFS_cross_sigmodel2}, we concentrate on the derivation of single-path scenarios as $R[n,l]=\sum_{i=1}^{N_{P}}R_{i}[n,l]$ and $r(t)=\sum_{i=1}^{N_{P}}r_{i}(t)$. By combining the continuous waveform $s(t)$ in \eqref{transmitted_waveform} and the output of multipath LTV channel in \eqref{Rx_baseband_signal_DSE}, \eqref{S2R_derive_process} can be derived with the help of \eqref{Tx_time_discrete} and $p_{i}=\frac{f_{c}}{\nu_{i}}=\frac{f_{c}N}{k_{i}\Delta{f}}$, which finishes the proof of Theorem \ref{th1_h_S2R}.
\section*{Acknowledgment}
This work was supported in part by Tsinghua University-China Mobile Research Institute Joint Innovation Center.

\bibliographystyle{IEEEtran}
\bibliography{ref-sum}

\end{document}